\documentclass[reprint,rsi,aip]{revtex4-1}

\usepackage{graphicx}

\begin{document}

\title{Active noise cancellation in a suspended interferometer}
\author{Jennifer C. Driggers}
\affiliation{LIGO Laboratory, California Institute of Technology, Pasadena, CA 91125}
\email{jenne@caltech.edu}

\author{Matthew Evans}
\affiliation{LIGO Laboratory, Massachusetts Institute of Technology, Cambridge, MA 02139}

\author{Keenan Pepper}
\affiliation{University of California, Berkeley, Berkeley, CA 94720}

\author{Rana Adhikari}
\affiliation{LIGO Laboratory, California Institute of Technology,
  Pasadena, CA 91125}

\date{\today}

\begin{abstract}
We demonstrate feed-forward vibration isolation on a suspended Fabry-Perot interferometer
using Wiener filtering and a variant of the common Least Mean Square (LMS) adaptive filter
algorithm. We compare the experimental results with theoretical estimates of the cancellation
efficiency. Using data from the recent LIGO Science Run, we also estimate the impact of this
technique on full scale gravitational wave interferometers. In the future, we expect to use
this technique to also remove acoustic, magnetic, and gravitational noise perturbations from
the LIGO interferometers. This noise cancellation technique is simple enough to implement in
standard laboratory environments and can be used to improve SNR for a variety of high precision
experiments.
\end{abstract}
\pacs{04.80.Nn, 95.55.Ym, 07.60.Ly, 42.62.Eh}
\maketitle

\section{Introduction}

The next generation of interferometers for gravitational-wave
detection, including the Laser Interferometer Gravitational Wave
Observatory (LIGO), will have 
unprecedented sensitivity to astrophysical events~\cite{RPP:S5}. At low 
frequencies ($\sim$10~Hz) it is likely that the displacement noise of the 
suspended mirrors will be limited at the 10$^{-20}$ $\mathrm{m} / \sqrt{\mathrm{Hz}}$ level by 
fluctuations in the Newtonian gravitational forces~\cite{Saulson:GGN,
  Thorne:GGN, VirgoNN}. 
The sources of the fluctuations are density perturbations in the environment
(e.g. seismic and acoustic) and mechanical vibrations of the nearby experimental 
apparatus. While care will be taken to mitigate the sources of all of these 
fluctuations, further reductions of this Newtonian noise may be made by carefully 
measuring the source terms and subtracting them from the data stream (offline) 
or in hardware by applying cancellation forces to the mirror.

To demonstrate the efficacy of this technique, we demonstrate below that offline 
subtraction of seismic noise can be done using static Wiener filtering based on 
an array of seismic sensors. The online adaptive subtraction is also shown to 
approach the 'optimal' Wiener limit~\cite{Haykin:Leaves}. Through
nonlinear processes, seismic noise below 10 Hz has been shown to limit
the performance of gravitational-wave detectors. This technique will
prove to be of substantial value in reducing the non-stationarity of
the detectors.

\section{Static Wiener Filtering}
\label{sec:StaticWiener}

To find a linear filter that will improve a chosen signal, we must first define 
what it means to 'improve' the signal. The figure of merit $(\xi)$
that we use for calculating the Wiener filters in this case is the expectation 
value of the square of the error signal ($\vec{e}$), where the error signal is 
defined as the difference between the target signal to be minimized and the estimate 
of that signal calculated from the filtered witness channels.

\begin{equation}
\xi \equiv E[e^2(n)] = E[d^2(n)] - 2 \vec{w}^\textrm{T} p + \vec{w}^\textrm{T} R \vec{w}
\label{eq:FOM}
\end{equation}

Here, $E[*]$ indicates the expectation value of $*$, $\vec{w}$ represents the tap 
weights of the filter, $d(n)$ is the 
target signal which we would like to minimize, $\vec{p}$ is the cross-correlation 
vector between the witness and target signals, and $R$ is the autocorrelation 
matrix for the witness channels.
When we solve Eq.~\ref{eq:FOM} by setting 
 
\begin{equation}
 \frac{d \xi}{d w_i} = 0
 \label{eq:optimize}
\end{equation}
 we find 
\begin{equation}
 R \vec{w}_{optimum} = \vec{p}
 \label{eq:Rwp}
\end{equation}

Eq.~\ref{eq:Rwp} finds the FIR (Finite Impulse Response) filter coefficients which minimize the RMS 
of the error $\vec{e}$ by optimizing the estimate of the transfer function 
between the witness sensors and the target signal.  Since the matrix
$R$ is Block Toeplitz, we take advantage of the
Levinson-Durbin~\cite{Durbin}
method of solving problems of the form $\vec{b} = M \vec{a}$, where $M$ is a
Toeplitz matrix. The Levinson method is considered weakly stable, as
it is susceptible to numerical round-off errors when the matrix is
close to degenerate (i.e. two or more witness sensors carry nearly
identical information about the noise source). For well
conditioned matricies it is much faster than brute force inversion of
the matrix~\cite{Huang:2006}.

The filtered output of the witness signals is 
$y(n) = \vec{w}^\textrm{T} \vec{x}$ and $e(n) = d(n) - y(n) = d(n) - \vec{w}^\textrm{T} \vec{x}$ 
is the filtered or minimized target signal, where $\vec{x}$ is the independent 
witness signal measuring our noise source. This new minimized 
signal $\vec{e}$ represents what the the disturbance $\vec{d}$ would
be if we were able to subtract the seismic noise before it entered the
system.

Since we are using this feed-forward technique to reduce noise in the
length of our cavities, and the cavity length carries gravitational
wave information, we must be careful not to subtract the
science signal along with the noise. For seismic noise, we are trying
to cancel noise below 10~Hz, while the gravitational wave signal band
above 10~Hz.  Even so, if we use a filter of the same length as
the data stream, we could perfectly cancel everything, including
seismic noise and gravitational wave information.  However given the short
length of our filters and the long stretches of data we calculate the
subtraction over, the distortion of the gravitational wave signals
should not be significant.
 
While the Wiener filtering occurs entirely in the time-domain, we examine plots 
in the frequency-domain of the filtered and unfiltered signals to determine 
the level of subtraction achieved. In Section~\ref{sec:Wiener40m} we
describe the Wiener simulations on a cavity in our lab. In
Section~\ref{sec:Wiener4km}, we make similar estimates for one of the 4~km LIGO
interferometers. Finally, in Section~\ref{sec:OAF} we demonstrate the
performance of a real-time seismic noise cancellation system in the lab.

\section{Wiener Filtering at the 40m Interferometer}
\label{sec:Wiener40m}
At our 40~m prototype interferometer~\cite{Rob:40m} lab at Caltech, both static 
Wiener filtering and adaptive filtering algorithms have been applied to a 
suspended Fabry-Perot triangular ring cavity's feedback signal.  We have used 
2~G\"{u}ralp CMG-40T seismometers~\cite{Guralp} and several Wilcoxon
731A accelerometers~\cite{Wilcoxon} 
as our independent witness channels ($\vec{x}$ in Section~\ref{sec:StaticWiener}),
and the low-frequency feedback signal for the cavity length as the target 
channel ($\vec{d}$ in Eq.~\ref{eq:FOM}) to reduce.  

\begin{figure}[htbp]
\centering
\includegraphics[width=.45\textwidth]{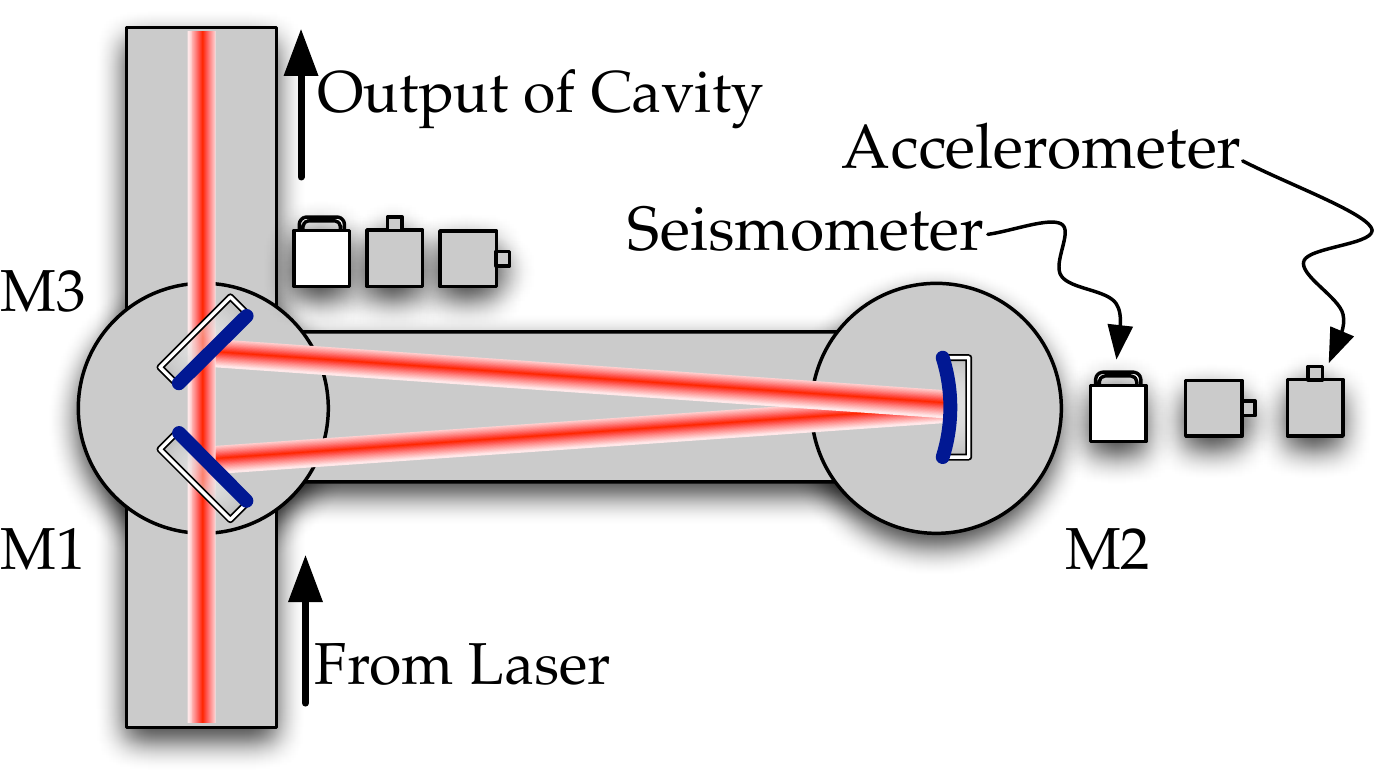}
\caption{Locations of seismometers and accelerometers in relation to
  the cavity mirrors. Round trip length of the triangular cavity is 27~m.}
\label{fig:MClayout}
\end{figure}

Figure~\ref{fig:MClayout} shows the locations of the witness sensors relative 
to the cavity mirrors.  The mirrors of the cavity are suspended as
pendulums with a resonance of $\sim$1~Hz
to mechanically filter high frequency noise, with the suspensions sitting on 
vibration isolation stacks to further isolate the optics from 
ground motion.  The 'stacks' are a set of 3 legs supporting the optical 
table on which the mirror sits, with each of the legs consisting of alternating 
layers of stainless steel masses and elastomer
springs~\cite{Giaime:Thesis, Giame:Stacks}.

\begin{figure}[htbp]
   \centering
    \includegraphics[width=0.5\textwidth]{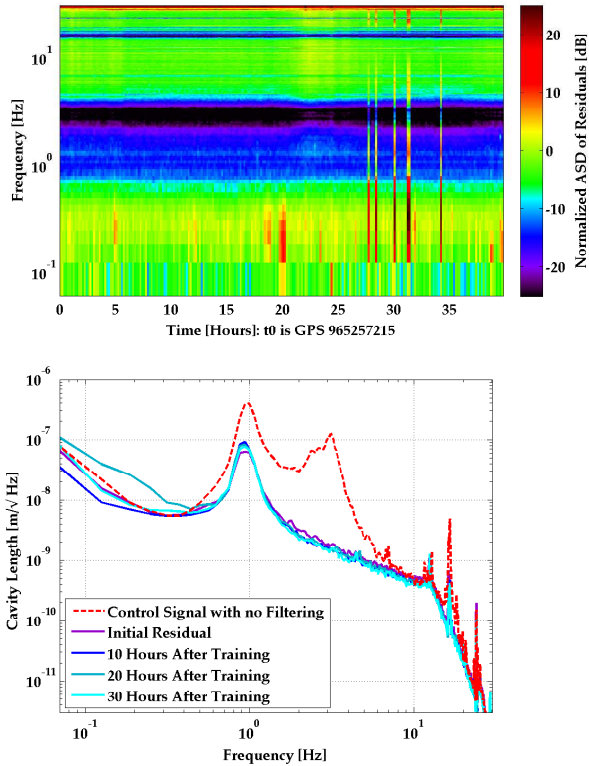}
   \caption{(Color online) Result of offline seismic Wiener filtering
     on suspended triangular cavity. (a) Spectrogram showing the efficacy of a
     Wiener filter applied offline over a several hour
     period. Noticeably different traces between $\sim$28~hours and
     $\sim$34~hours are the result of non-stationary anthropogenic noise, not a
     decay of the filter's efficacy. (b)
     Amplitude spectral density of the control signal. Dotted red is without
     subtraction, purple is initial residual, progressively lighter
   blues are 10 hours, 20 hours and 30 hours after filter was trained.}
   \label{fig:40mMCsurf}
\end{figure}

We used Matlab~\cite{mathworks} to import the data for the length feedback signal for our cavity, 
and to construct and apply the Wiener filters. Since the feedback
control bandwidth is $\simeq$50~Hz, the feedback signal can be used as
an accurate measure of the seismic disturbance at low frequencies.
In 
Figure~\ref{fig:40mMCsurf} we show results of a day-long simulation study.  This study was 
done to determine the length of time we can use a set of static filters 
before updating. We use
1~hour of data to train and calculate a single Wiener filter, and then 
apply that filter to 10~minute segments of data for one day, using a
31 second long,
2000 tap filter with a sample rate of 64~Hz.  In
Figure~\ref{fig:40mMCsurf}b, we 
select a few typical 
traces to illustrate the capabilities of the filter, while in 
Figure~\ref{fig:40mMCsurf}a we show the full results as a spectrogram,
whitened by normalizing to the spectra during the time the filter was
being trained.  We see large amounts of noise reduction both 
at the broad stack peak at $\sim$3~Hz and around the 16~Hz vertical mode 
of the mirror pendula.  

\begin{figure}[htbp]
  \centering
  \includegraphics[width=0.5\textwidth]{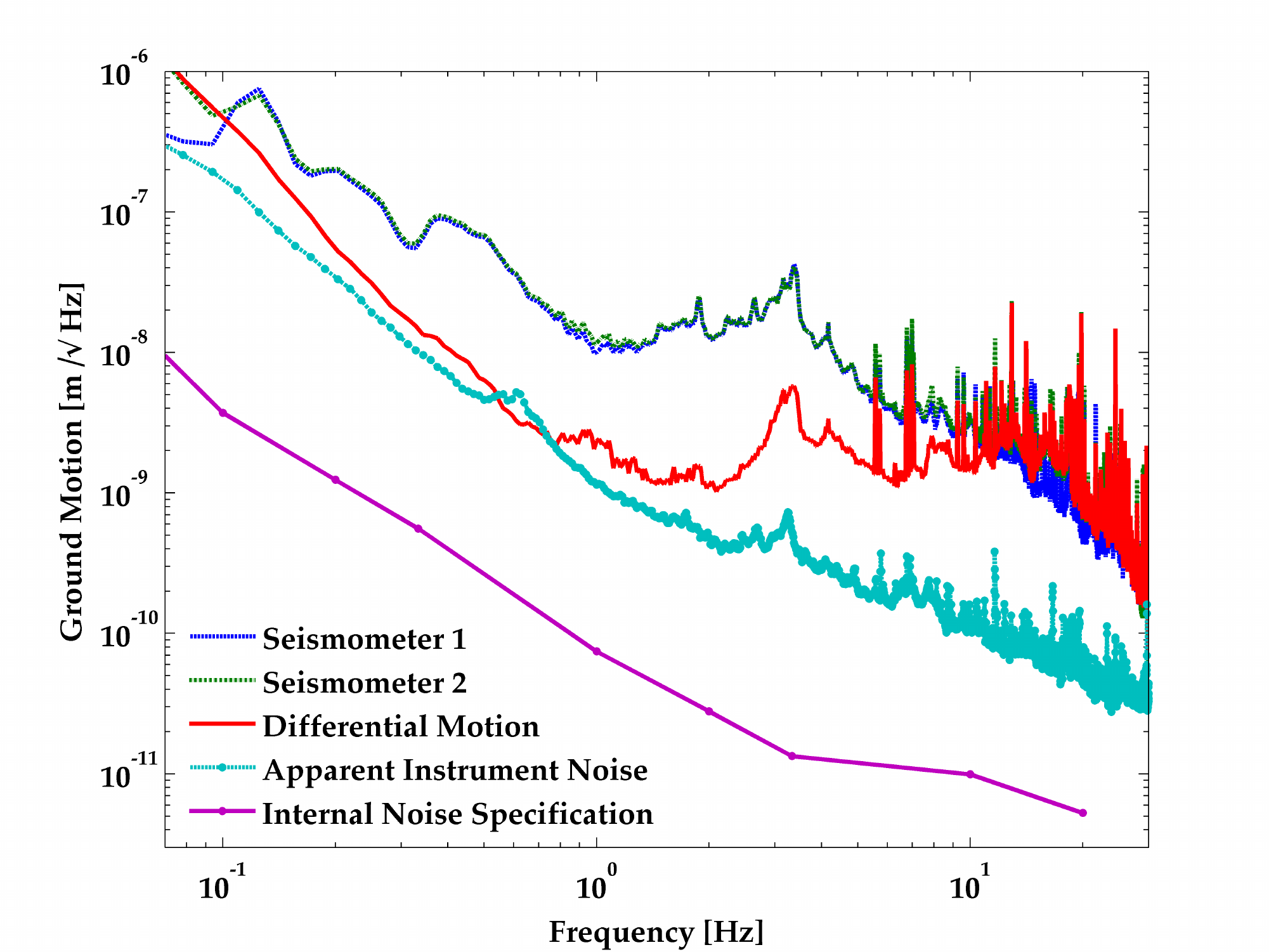}
  \caption{(Color online)  Shown are the spectra of the individual
    seismometers (blue dashed and green dash-dot),
    the manufacturer's spec for the seismometers' internal noise
    (purple solid-circle), and the
    differential ground motion along the 13.5 m length of the
    cavity (red solid). 
    We also show the differential noise of the seismometers with the seismometers
    collocated in a stiff seismic vault (teal dash-circle); in principle, this is a measurement of the
    actual seismometer noise floor. It is unknown what uncorrelated
    noise is present in our sensors which makes the teal trace so much larger
    than the specification.}
  \label{fig:MCDiffMotion}
\end{figure}

%\begin{figure}[htbp]
  % \centering
  % \includegraphics[width=.5\textwidth]{./Figures/Rana-elog-WhitenedMCLresidualafterMISOwiener.png} 
  % \caption{Relative fluctuations in the residual shown in Figure \ref{fig:40mMCspaghetti} over 5 hours. 
  % LABEL THE Z AXIS AND USE dB. REDO ALONG WITH SPAGHETTI PLOT: LONGER DURATION}
 %  \label{fig:40mMCsurf}
%\end{figure}

We also include the noise contributions of our seismometers in Figure~\ref{fig:MCDiffMotion} to
demonstrate how close we are able to get to the fundamental limit of Wiener filtering.  
Since the Wiener filter accepts, as inputs, the signals from the witness sensors (which have 
true ground motion information plus self-noise of the instruments and noise in the readout 
electronics), all of these noise contributions are filtered and added 
back into our data stream, limiting our ability to suppress ground motion below these levels.
In Figure~\ref{fig:MCDiffMotion}, we show
that the differential ground motion over the length of the cavity is not much larger than
the instrument noise of the seismometers. In other words, the ground
noise over length of the cavity is strongly correlated below $\sim$~1~Hz and so
the differential motion is much smaller than the motion of any
individual sensor.  Currently our measurement of the differential
ground motion is limited by the apparent instrument noise of the
seismometers, represented by the teal trace in
Figure~\ref{fig:MCDiffMotion}.  The apparent instrument noise is
significantly higher than the specification, which indicates that
there is some unknown noise which is uncorrelated between 2
seismometers, even when they are placed very close together.  
We will use lower noise sensors and readout electronics and better
thermal/acoustic isolation of the seismometers in order to get better performance on such short
baselines.

% \begin{figure}[htbp]
%   \centering
%   \includegraphics[width=.5\textwidth]{Figures/SeismoHuddle.pdf} 
%   \caption{
%            Seismometer noise estimated by enclosing the two units in
%            a foam box and on a granite slab. The Residual trace is
%            made by subtracting the coherent common mode signal between
%            the seismometers. Also shown is the instruments noise spec.}
%   \label{fig:HowLow}
%\end{figure}

The limit to the performance of the feed-forward subtraction seems to be a combination
of low frequency noise in the seismometers and the feedthrough of noise from the auxiliary
controls systems of the cavity (e.g. angular controls, pendulum damping servos, etc.).

%\clearpage
%\FloatBarrier
\section{Applying Wiener Filtering to a 4 km LIGO Interferometer}
\label{sec:Wiener4km} 
One of the LIGO sites in Livingston, Lousisiana has had a hydraulic
external pre-isolator (HEPI) actuation system installed since 2004
(the other LIGO site in Hanford, Washington will receive a HEPI system
as part of the Advanced LIGO upgrade) \cite{Norna:HEPI}.  This HEPI system is designed
to actuate on the seismic isolation stacks which support the suspended
LIGO optics to actively reduce seismic noise.  Initial implementation
of the HEPI actuators only included local seismic isolation between 0.1-5~Hz to reduce
anthropogenic noise, tidal effects and the microseism \cite{HEPIthesis}.

To estimate how the global Wiener filtering technique should scale up
to a full size interferometer, we 
analyzed data from the 5$^{\rm th}$ LIGO Science
Run~\cite{RPP:S5}. While this analysis was done as offline
post-processing, results from later tests executed on the LIGO
interferometers using the HEPI actuators during the 6$^{\rm th}$ LIGO Science
Run will be available in a future paper \cite{FFWpaper}. 

\begin{figure}[htbp]
   \centering
   \includegraphics[width=.5\textwidth]{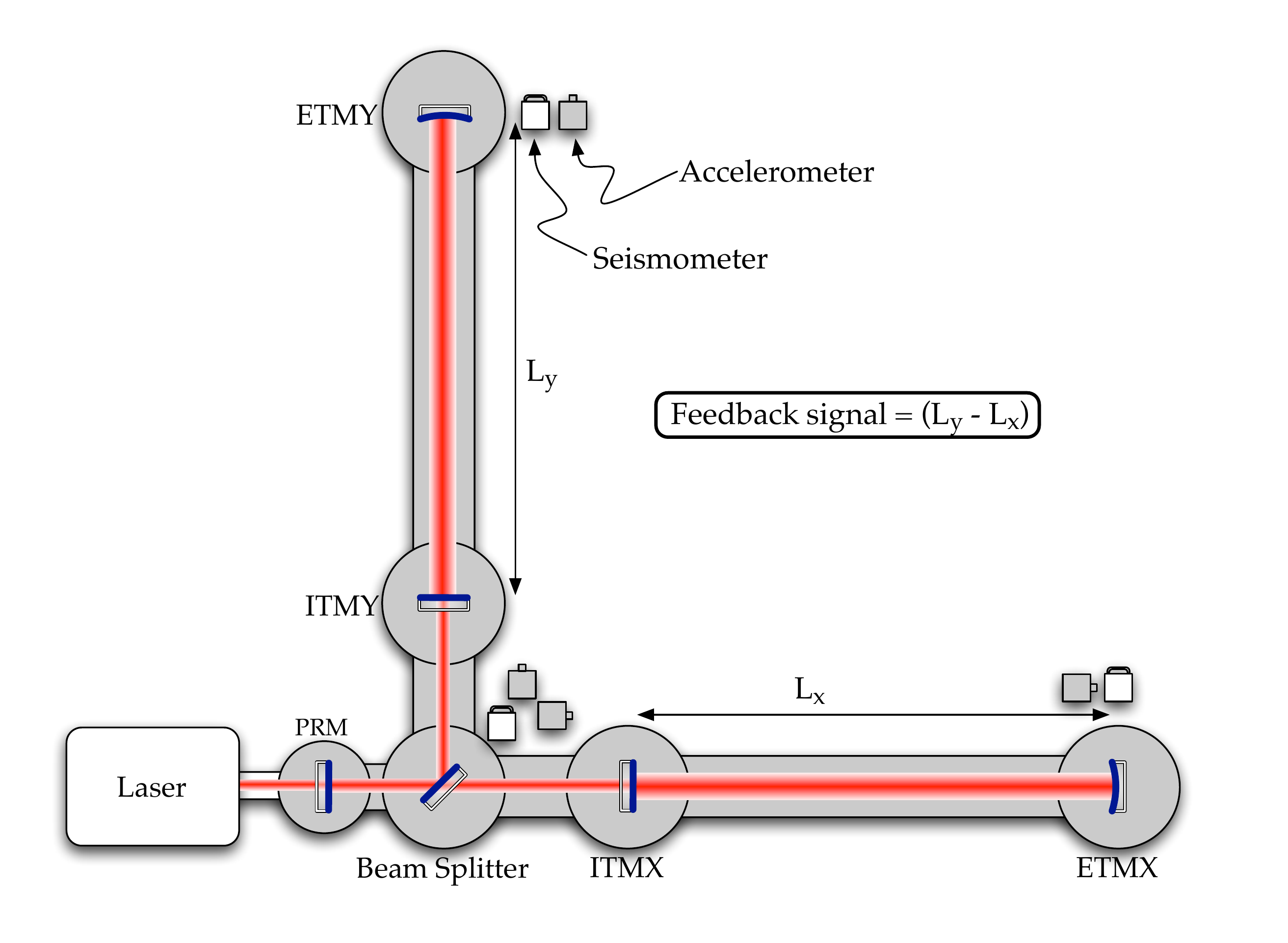} 
   \caption{Schematic layout of seismometers relative to interferometer mirrors.}
   \label{fig:H1layout}
\end{figure}

Instead of a single 
cavity, in this case we explored the subtraction of seismic noise from the differential 
arm length feedback signal (which is an accurate measure of the low frequency 
ground noise). The sensors are placed close to the ends of the interferometer arms 
and at the beamsplitter as shown in Figure~\ref{fig:H1layout}.

\begin{figure}[htbp!]
   \centering
   \includegraphics[width=.5\textwidth]{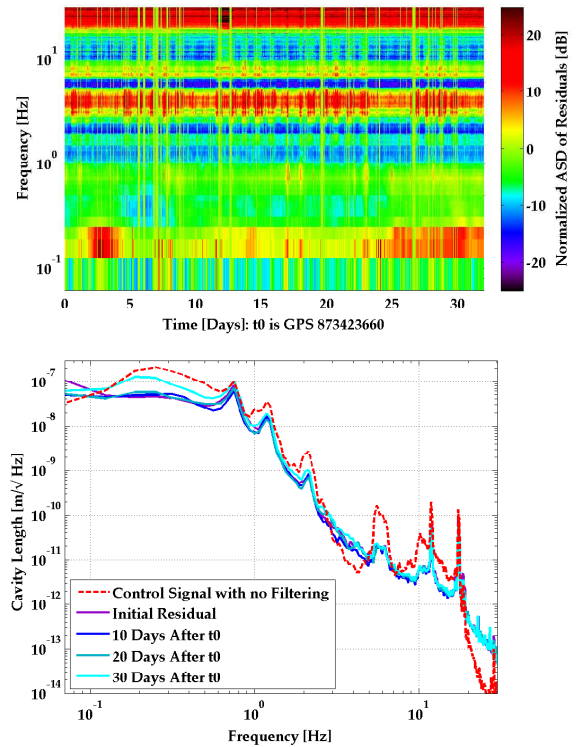}
   \caption{(Color online) Result of offline simulated seismic Wiener filtering on the 4km LIGO 
            Hanford interferometer. (a) Traces are amplitude spectra
            normalized to the unfiltered control signal (red trace in b), which is at a time during the filter's training. 
            Filter was trained on 6 hours of data, then applied in 10 
            minute segments.  Vertical stripes indicate times when the 
            interferometer was not operational.  Seismic subtraction
            is fairly constant on a one month time scale,
            although it is not particularly effective for times when seismic
            noise is significantly different from the training time (b) Selected individual
            spectra from a above.  Dotted red trace is before subtraction,
           purple trace is initial residual and progressively
           lighter blues are 10, 20 and 30 days after the filter was trained.}
   \label{fig:H1S5OneMonthSurf_NoWeight}
\end{figure}

Figure~\ref{fig:H1S5OneMonthSurf_NoWeight} shows the resulting
subtraction efficacy for a static filter. The
variation in the 0.1-0.3~Hz band comes from variation in the ambient level of
the double frequency microseismic peak~\cite{microseism1969}. The structure in the 1-15~Hz band is the
usual increase in anthropogenic noise during the workday.

Even though some excess noise is added in the dips around 3-5~Hz and 7~Hz, the
filter reduces the main contributors to the RMS of the control
signal, and the reduction is remarkably stable over the 30 day
timespan.  This static filter does inject an unacceptable amount of
noise above 20~Hz, which we will elimiate in the future by using more
aggresive pre-weighting to
disallow such noise amplification before calculating the Wiener filter.  

\begin{figure}[htbp!]
   \centering
   \includegraphics[width=.5\textwidth]{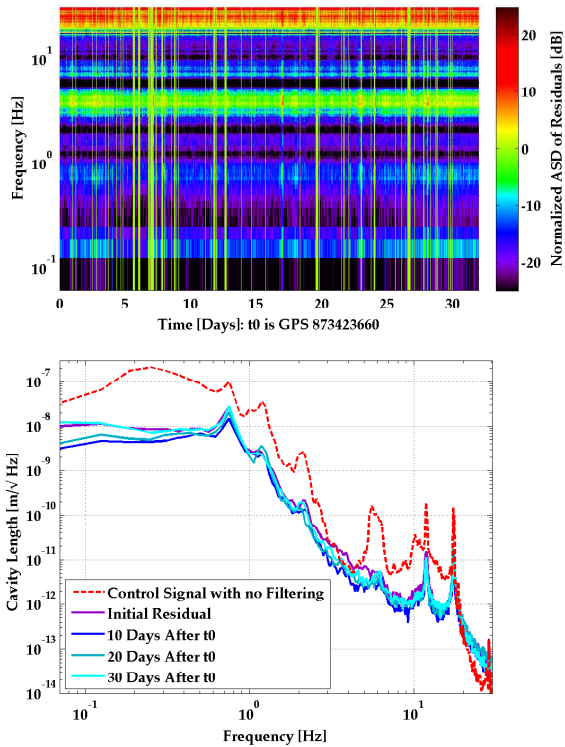}
   \caption{(Color online) Result of offline simulated seismic Wiener
     filtering on the 4km LIGO Hanford interferometer, using an
     acausal filter on the same 30 day data set.  (a) Traces are amplitude
     spectra normalized to the unfiltered control signal (red trace in
     b).  A filter is trained on,
   and then applied to, 10 minute segments of data.  Seismic noise is
   more effectively suppressed using this constantly updated filter,
   implying that the transfer function is changing on a relatively
   short time scale, and that it is advantageous to update the filter
   more often than once per month. (b) Selected individual 
spectra from a above.  Dotted red trace is before subtraction, 
purple trace is initial residual and progressively 
lighter blues are 10, 20 and 30 days after beginning. }
   \label{fig:H1S5OneMonthSurf_WithWeight}
\end{figure}

Figure~\ref{fig:H1S5OneMonthSurf_WithWeight} shows the subtraction if
we use an acausal filter, retraining it every 10~minutes, for the same
30~day data set.  This filter
performs much better than the static version.  While we cannot apply
an acausal filter in realtime, we can utilize causal adaptive filters to
achieve nearly the same effect as long as the seismic environment does
not change appreciably on time scales less than 10 minutes.

Residuals for both Figures~\ref{fig:H1S5OneMonthSurf_NoWeight}
and~\ref{fig:H1S5OneMonthSurf_WithWeight} were calculated using
46 second long Wiener
filters of 3000 taps at a sample rate of 64 Hz.

%\begin{figure}[htbp!]
%   \centering
%   \includegraphics[width=.5\textwidth]{./Figures/SpagHist_temp.pdf} 
%   \caption{individual spectra from Figure \ref{fig:H1S5OneMonthSurf}}
%   \label{fig:H1S5OneMonthSpaghetti}
%\end{figure}

\section{Online Adaptive Filtering at the 40m Interferometer}
\label{sec:OAF}
In case the transfer functions between the sensors and the target are changing 
with time, it would be useful to use a filter whose coefficients change with 
time. Such an adaptive filter could also take into account changes in the 'actuator'.
The most simple and common implementation of an adaptive filter is the 
Least Mean Squares (LMS) algorithm~\cite{Sayed:2003}.

The Online Adaptive Filtering (OAF) algorithm implemented at the 40m Lab is the 
Filtered-x Least Mean Squared (FxLMS) algorithm~\cite{Widrow:ASP}.  It is based 
primarily on the canonical LMS algorithm; a steepest descent optimization 
of a defined error function.  Just as in the static Wiener filtering in 
Section~\ref{sec:StaticWiener}, we minimize the RMS of the difference between 
the filtered output and the original feedback signal. The LMS algorithm described in 
Equation~\ref{eq:adaptive} takes 'steps' in the direction of the steepest gradient 
until it arrives at a local minima.  

\begin{equation}
w(n+1) = w(n) \times [1 - \tau] + \mu  \times e(n) \times  x(n)
\label{eq:adaptive}
\end{equation}

Here the next iteration's FIR coefficients depend on the the current 
coefficients ($w$), the current witness signal ($x$), the current error signal 
(difference between the target and filtered signal, $e$), and the adaptation rate 
($\mu$).  One of the largest challenges with the adaptive filtering algorithm is 
that the success of the algorithm is fairly sensitive to the choice of
$\mu$.  To improve stability against transients, we modify the usual FxLMS 
algorithm to include a decay constant $\tau$.

\begin{figure}[htbp]
   \centering
   \includegraphics[width=.48\textwidth]{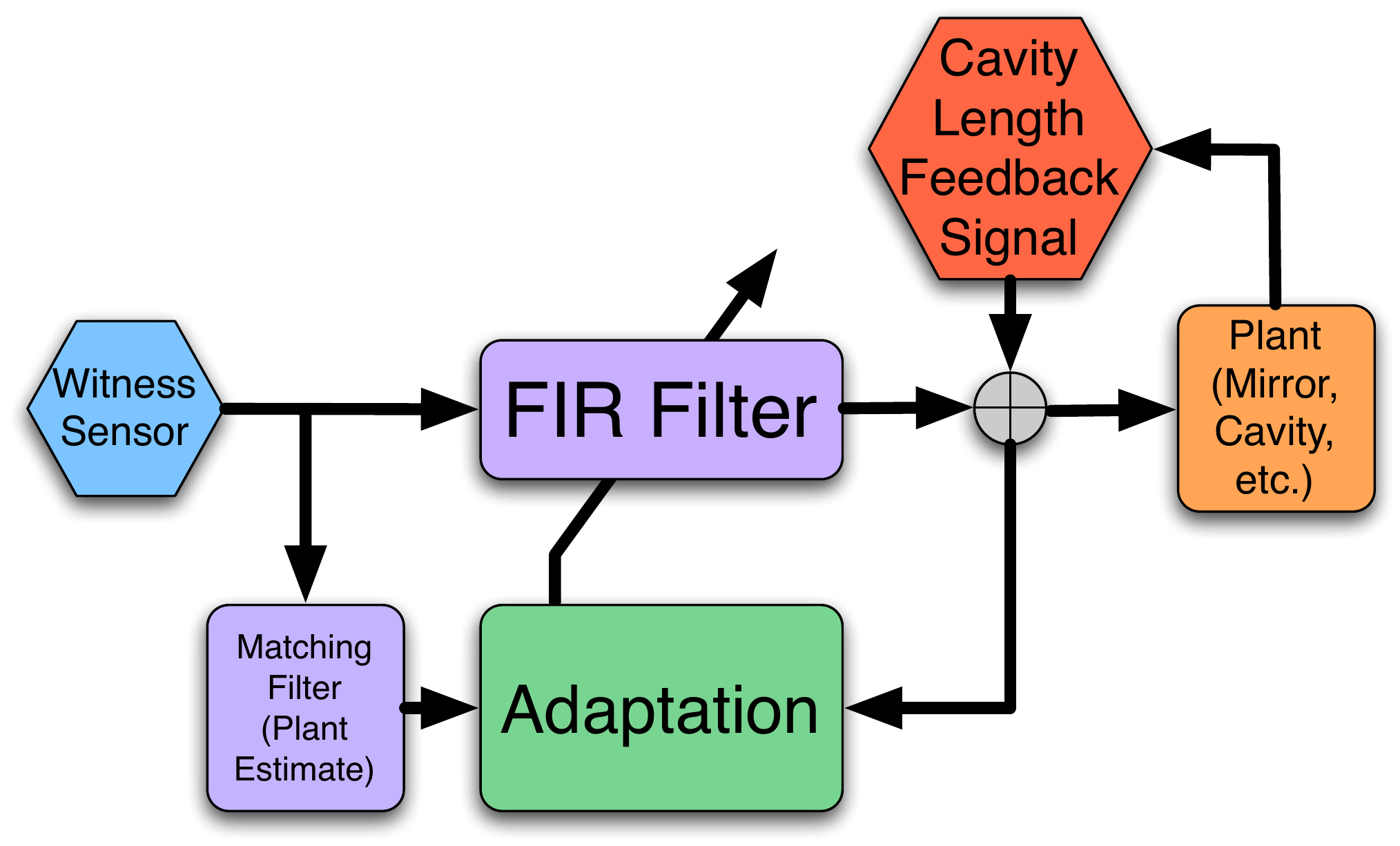} 
   \caption{Block diagram of the FxLMS algorithm used.}
   \label{fig:fxlms}
\end{figure}

The FxLMS algorithm acknowledges that there exist phase delays in the path 
of the target signal which cannot be approximated by the LMS method 
alone~\cite{NI:FXLMS}.  To account for these phase delays, we filter the incoming 
witness signals with filters identical to those in the target signal path.  Once we 
have matched the delays in the two different paths, we implement the regular LMS 
optimization to find the coefficients we will use in our FIR filter.  The FxLMS 
algorithm used is sketched in Figure~\ref{fig:fxlms}.

\begin{figure}[h]
   \centering
   \includegraphics[width=0.5\textwidth]{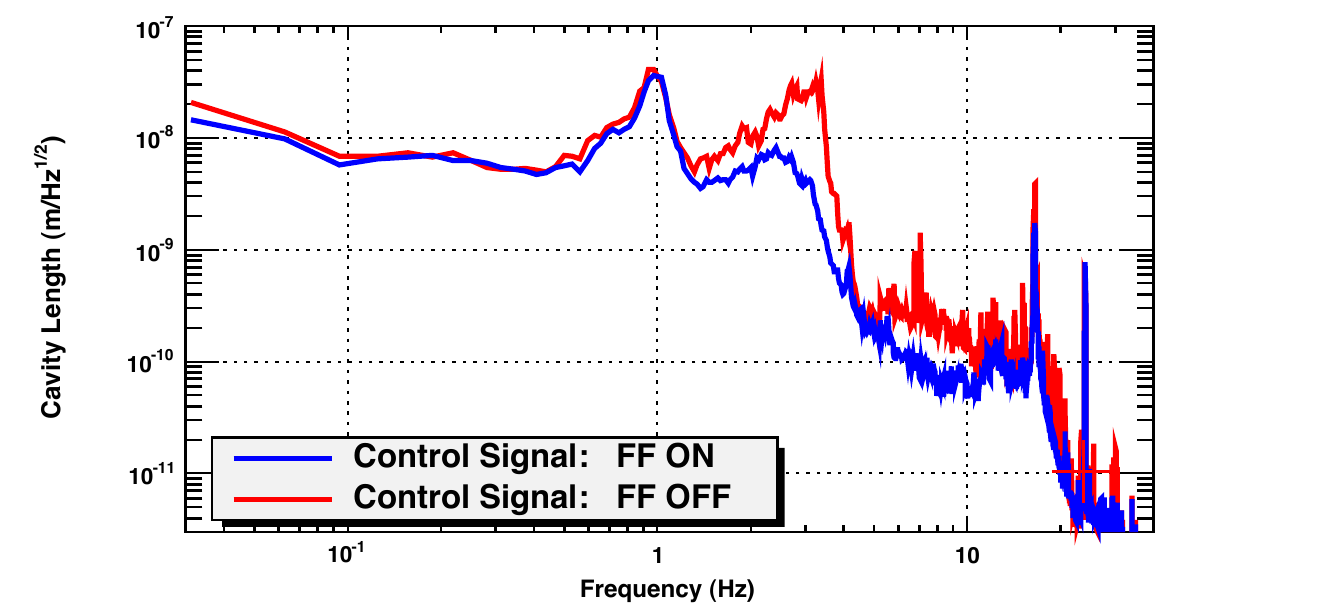} 
   \caption{(Color online) Online Adaptive Filter performance: the spectral density of the cavity length 
                  fluctuations are shown with the feed-forward on
                  (lower blue trace) and off (upper red trace).}
   \label{fig:40mMCadaptive}
\end{figure}

We apply the OAF system to the same triangular cavity as in 
Section~\ref{sec:Wiener40m}. Once again, we use the cavity length feedback 
signal as our targeted signal to minimize, and a similar layout of independent witness 
sensors as shown in Figure~\ref{fig:MClayout}.  Unlike Sections \ref{sec:Wiener40m}
and \ref{sec:Wiener4km} which were simulations using previously collected data, 
here we are actuating on the cavity in realtime.
Figure~\ref{fig:40mMCadaptive} shows results using a 125 second long, 2000 tap filter
with $\mu=0.01$ and $\tau=10^{-6}$ at a 16 Hz sample rate.  The on /off traces
in the adaptive case are similar to estimates made in the static Wiener
filtering case (Figure~\ref{fig:40mMCsurf}).  Given enough time to
adapt, the OAF converges towards the optimal filter, but, so far, not
completely.  
Since the adaptive system was tested using one G\"{u}ralp seismometer~\cite{Guralp} and
one Ranger SS-1 seismometer~\cite{Ranger}, the subtraction is not as
pronounced as if we had used 2 G\"{u}ralps, or other more sensitive
broadband seismometers. In the next iteration of this setup, we will
explore the variation in the cancellation performance as a function of
sensor placement.

% Good idea - we should always keep the .xml source in the SVN directory
% and comment its location like this:
% This figure comes from .../Templates/OAF/OAF-100205.xml

\section{Conclusions}
We have demonstrated the use of Wiener filter based feed-forward seismic noise 
reduction on a suspended interferometer. We have also implemented a stable, adaptive 
feed-forward system which has a performance approaching that of the
optimal Wiener estimate. These techniques can be simply implemented in any general laboratory 
requiring vibration isolation using relatively low cost accelerometers and commodity 
computers and DSP software (e.g. Labview). These 'optimal' feed-forward schemes 
operate without having to know \textit{a priori} the transfer function between the 
disturbance and the primary experiment; they can easily be reconfigured to adapt 
to new experimental setups.  Similar Wiener filter and LMS based
techniques have been utilized in other experiments, both offline and
online, for example isolating 2-mirror Fabry-Perot cavities from ground
motion~\cite{FPphasenoise}, reducing acoustic noise in oceanography 
settings~\cite{oceanographyAcoustics} and in signal processing to
decorrelate degenerate witness channels~\cite{GramSchmidtDecorrelation}.

In the near future, we will work to use this scheme to reduce the noise in 
multiple degrees of freedom of the full interferometer. It is clear that this 
technique can also be applied to remove other sources of environmental noise 
(e.g. acoustic, magnetic, electronic, etc.). For each noise source in
the gravitational wave band, we will inject software
gravitational wave signals into the data stream and confirm that they are not distorted.  There will certainly be new challenges 
associated with each type of noise, but this seems promising as a method which can be employed 
to reduce the influence of environmental noise in a wide variety of experimental 
setups.

\section{Acknowledgements}
We gratefully acknowledge illuminating discussions with Joe Giaime, Alan Weinstein, 
Rob Ward, and Jan Harms. We also thank the National Science Foundation for support 
under grant PHY-0555406. J. Driggers also acknowledges the support of an NSF
Graduate Research Fellowship. K. Pepper acknowledges the support of
the LIGO NSF REU program.  LIGO was constructed by the California Institute of 
Technology and Massachusetts Institute of Technology with funding from the National 
Science Foundation and operates under cooperative agreement PHY-0107417. This article 
has LIGO Document Number P0900071.

% Use this one if working on the bib.  Otherwise copy in the .bbl file!!!
%\bibliography{AFbib_noTitles}
%

%%%%% Some errors are because BibDesk can't handle the long author
%%%%% lists.  Compile with just ~1 author, then hand-copy in the other
%%%%% authors.  ---JD 12Sept 2011

%merlin.mbs 2010-03-15 4.21a (PWD, AO, DPC)

\end{document}